# Proposal for a Negative Capacitance Topological Quantum Field-Effect Transistor

M.S. Fuhrer[1,2], M.T. Edmonds[1,2], D. Culcer[1,3], M. Nadeem[1,4], X. Wang[1,4], N. Medhekar[1,5], Y. Yin[1,2,5], J.H Cole[1,6]
[1]ARC Centre of Excellence in Future Low-Energy Electronics Technologies (fleet.org.au), [2]School of Physics and Astronomy, Monash University, 3800 Victoria, Australia, email: michael.fuhrer@monash.edu, [3]School of Physics, University of New South Wales, Sydney 2052, Australia, [4]Institute for Superconducting and Electronic Materials (ISEM), Australian Institute for Innovative Materials (AIIM), University of Wollongong, Wollongong, New South Wales 2525, Australia, [5]Department of Materials Science and Engineering, Monash University, Clayton VIC 3800, Australia, [6]Chemical and Quantum Physics, School of Science, RMIT University, Melbourne, VIC 3001, Australia.

*Abstract*—A topological quantum field effect transistor (TQFET) uses electric field to switch a material from topological insulator ("on", with conducting edge states) to a conventional insulator ("off"), and can have low subthreshold swing due to strong Rashba spin-orbit interaction. Numerous materials have been proposed, and electric field switching has been demonstrated in ultrathin $Na_3Bi$. Here we propose a negative capacitance (NC) TQFET which uses a ferroelectric to amplify the electric field and potentially achieve very low switching voltages and energies. Materials challenges for realizing the NC-TQFET are discussed.

## I. Electric Field Switching of Topology

Electric field effect switching from topological insulator (TI) to conventional insulator has been proposed in a number of materials [1], and has been demonstrated in ultrathin (monolayer or bilayer) $Na_3Bi$ (Figs. 1-3) [2].

## II. Reducing Subthreshold Swing with Rashba Spin-Orbit Interaction

The operation of a transistor in low-power applications such as switches is quantified by the sub-threshold swing $S = [dln(I)/dV_g]^{-1}$ where $I$ is the current, and $V_g$ the gate voltage. In a conventional transistor $S$ is limited by Boltzmann's tyranny to 60 mV/decade when the gate is perfectly coupled to the channel. In a topological transistor, a gate field induces a sublattice potential difference $\lambda_v$ and opens a gap, inducing a topological phase transition between insulating and conducting states, which replaces conventional carrier inversion (Fig. 4). The effective sub-threshold swing $S^*$ is given by

$$S^* = e \left[\frac{dE_g}{d\lambda_v}\right]^{-1} \quad (1)$$

such that $S^* = 1$ corresponds to Boltzmann's limit. In the simplest models the gap equates to the potential difference established by the gates $\lambda_v$, and one expects $S^* = 1$. Yet certain materials, such as honeycomb Xene lattices of heavy atoms, experience a strong Rashba spin-orbit interaction due to the gate electric field (Fig. 5). As a result, $S^*$ can be less than 1 (Fig. 6). We have termed this the TQFET (Fig. 7a-f). We found $S^* < 0.75$ in existing materials and estimate $S^*$ as low as 0.57 in functionalized Bi (Table 1) [3].

## III. Unipolar, Single-Gate TQFET

As described in Ref. [3] the TQFET is ambipolar, turned off by either $V_b > 0$, $V_t < 0$ or $V_b < 0$, $V_t > 0$. Ideally an FET device should operate with a single gate as a unipolar transistor. This situation is realized by grounding one gate and applying a gate voltage to the other. Furthermore if the NC-TQFET channel is connected to semiconducting source/drain leads, unipolar conduction results (Fig. 8). At $V_g = 0$ the electric field is zero and the device is "on". At $V_g < 0$ the electric field is non-zero, opening a bandgap, and the net effect of the two gates is to shift the overall potential by and amount $V_g/2$. This allows a bandgap $eV/S^*$ to be opened at a gate voltage $V$.

## IV. Proposal for a Negative Capacitance TQFET

The TQFET is a true electric-field effect device, where the barrier to conduction is created through an electric field, rather than raising an existing potential barrier in the channel, and thus can take full advantage of electric field amplification by e.g. a negative capacitance [4], whereas in an ideal conventional FET the electric field is zero in the channel in the subthreshold region[5]. In the NC-TQFET a 2D TI layer is sandwiched between two ferroelectric layers, with independent gates on top and bottom (Fig. 9a). The combined structure of ferroelectric (negative capacitance) and 2D TI (positive capacitance) amplifies the electric field in the 2D TI layer (Fig. 9b), similar to the negative capacitance FET proposed in Ref. [4].

The NC-TQFET is in the "on" state when the two top gate voltages $V_b = V_t = 0$; the potential across the ferroelectric/2D TI/ferroelectric structure is constant, the electric field in the 2D TI is zero, and the 2D TI is in a topological state (red circle in Fig. 9c). Applying gate voltages $V_b > 0$, $V_t < 0$ produces the potential profile shown in blue in Fig. 9b, amplifying the electric field in the 2D TI, and opening a conventional bandgap ("off" state; blue circle in Fig. 9c).

We now estimate the electric field amplification. We can approximate the electric field-polarization $E$-$P$ relationship for the ferroelectric layer as

$$E = 2\alpha_{FE}P + 4\beta_{FE}P^3 \quad (2)$$

Where $\alpha_{FE} < 0$ and $\beta_{FE} > 0$ are parameters that characterize the ferroelectric. A negative value of α indicates a bistable $P(E)$ relationship i.e. ferroelectricity. The TI layer has $E = 2\alpha_{TI}P$ with $\alpha_{TI} = +1/(2\varepsilon)$ corresponding to a linear dielectric. Following Ref.[4], we find a relationship between the gate voltage applied across the FE/TI/FE stack, $V_g$, and the surface potential difference across the TI, $\psi_s$:

$$V_g = (1 + a_1)\psi_s + a_2\psi_s^3 \qquad (3)$$

where $a_1 = \alpha_{FE}/\alpha_{TI}$ and $a_2 = (4\beta_{FE}/\alpha_{TI}^3)t_{FE}^2$ where $t_{FE}$ is the ferroelectric total thickness (twice the thickness of top and bottom layer). We will have the maximum electric field amplification when $\alpha_{FE}/\alpha_{TI} = -1$ or $t_{FE} = (2|\alpha_{FE}|C_{TI})^{-1}$. Then

$$V_g = \frac{2\beta_{FE}C_{TI}^2}{|\alpha_{FE}|}\psi_s^3 = \frac{C_{TI}^2}{P_r^2}\psi_s^3 \qquad (4)$$

where $P_r = \sqrt{-\alpha/2\beta}$ is the remanent polarization.

## V. Estimating NC-TQFET Performance

We first consider bilayer graphene (BLG) as a material with an experimentally characterized electric-field-dependent bandgap [8,9]. A family of silicon-compatible ferroelectrics based on $HfO_2/ZrO_2$ has been identified; we assume $P_r = 27.5$ μC/cm$^2$ for our NC-TQFET corresponding to La-doped $HfO_2$[6].

For a material with negligible spin-orbit coupling such as bilayer graphene, we expect $S^* = 1$. However, Ref. [3] neglected screening by the active layer. The electric field is reduced by dielectric screening in the BLG layer by a dielectric constant $\kappa$. Furthermore, the separation of the atoms in the sublattice $t_v$ is smaller than the van der Waals thickness of the layer $t_{TI}$, which further reduces the sublattice potential difference by an amount $t_v/t_{TI}$, such that $\lambda_v = (t_v/t_{TI}\kappa)\psi_s$, and $dE_g/d\psi_s = et_v/t_{TI}\kappa S^*$. For BLG, $t_v/t_{TI} = 0.5$, $\kappa = 3.6$ [7], $S^* = 1$ predicts $dE_g/d\psi_s = 0.139$ which is very close to experimentally measured values from optical spectroscopy [8] and electronic transport [9]. The BLG capacitance $C_s = 0.048$ F/m$^2$ (assuming $\kappa = 3.6$, $t_{TI} = 6.68$ Å) allows $E_g(V_g)$ to be calculated for the ferroelectric/BLG/ferroelectric structure (solid black curve, Fig. 10). If we assume a *hypothetical* NC-TQFET which is a strongly spin-orbit coupled version of BLG with similar screening properties but $S^* = 0.57$, equivalent to the spin-orbit parameters for Bi from Ref. [3], we have $E_g(V_g)$ given by the solid blue curve in Fig. 10.

We compare the hypothetical NC-TQFET a hypothetical low voltage CMOS device (CMOS LV) which corresponds to a GAAFET at the 2018 node (metal-1 half-pitch $F = 15$ nm) [10]. CMOS LV characteristics are $V_g = 0.3$ V, on current $I_{on} = 3.2$ μA, on-off ratio $I_{on}/I_{off} = 1.3 \times 10^4$, intrinsic device switching energy $E_{int} = 3.62$ aJ and intrinsic delay time $\tau = 3.8$ ps. We assume that the off current is $I_{off} = I_{on}\exp(-E_g/kT)$ where the $I_{on}$ is the current at $E_g = 0$, $kT = 26$ meV is the thermal energy at room temperature, which corresponds to $E_g = 245$ meV. From Fig. 10 we see BLG has $E_g = 245$ meV at $V_g = 0.213$ V, only a modest improvement over CMOS LV. However, the hypothetical NC-TQFET (blue curve) has $E_g = 245$ meV at $V_g = 0.030$ V. This represents an order-of-magnitude improvement over CMOS LV.

We further assume that the NCTQFET has a width $F$ and gate length $F = 15$ nm. The on-state occurs at zero bandgap, where the BLG is a massive Dirac semimetal with a conductivity approximately 4 $e^2/h$ and conductance 155 μS. Drain voltage $V_d = V_g = 0.030$ V gives $I_{on} = 4.6$ μA, similar to CMOS LV. The gate charge is $6.0 \times 10^{-17}$ C = 374 $e$, and the intrinsic switching energy $E_{int} = (1/4)QV_g = 0.45$ aJ. The channel resistance $R = 6.45$ kOhms, and gate capacitance $C = Q/V_g = 2000$ aF, giving $\tau = (RC) = 13$ ps.

The intrinsic switching energy of 0.45 aJ is almost an order of magnitude lower than CMOS LV and lower than all devices considered in [10] except the BISFET [11] and related ITFET. However, we have made a number of idealistic assumptions. First and most important, there is no known material with the properties of the NC-TQFET channel we consider here. $S^* = 0.57$ was estimated for a functionalized staggered sublattice of bismuthene in Ref. [3]. However, we might expect that $t_v/t_{TI}$ is small for such a lattice due to the extra thickness of the functionalization layer, and the dielectric constant $\kappa$ is not known. Second, our assumptions about the device structure are optimistic. We assume that the capacitance of the structure is described by an ideal ferroelectric described by Eqn. (1) in series with the linear dielectric of the TI layer which is certainly a gross approximation, and that $\alpha_{FE}/\alpha_{TI} = -1$ can be achieved perfectly.

Nevertheless, the NC-TQFET points to a general strategy to realize a new type of low-voltage transistor. The operating parameters of such a transistor are set by the materials parameters of the 2D TI and ferroelectric layers, and there appears to be no fundamental lower bound to the subthreshold swing for such a device. In particular, there appears to be no fundamental reason why a 2D TI cannot be designed with suitably low $S^*$, high $t_v/t_{TI}$ and low $\kappa$. Further work, guided by first-principles prediction of new materials, is needed to identify candidate 2D TI materials for NC-TQFETs. The ferroelectric may also be optimized; increasing the remanent polarization of the ferroelectric would reduce the operating voltage $V_g \sim P_r^{-2}$. For example, remanent polarization >130 μC/cm$^2$ has been observed in tetragonal-like $BiFeO_3$ [12], which would correspond to a >20-fold reduction in the operating voltage (dotted curves, Fig. 10).

### Acknowledgment

The authors acknowledge support from the ARC Centre of Excellence in Future Low-Energy Electronics Technologies.

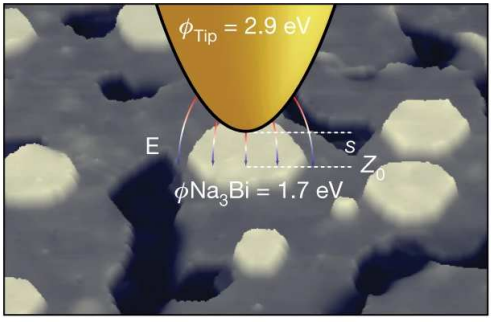

Fig. 1. Schematic of measurement of the electric field dependence of the bandgap in Na$_3$Bi Adapted from [2].

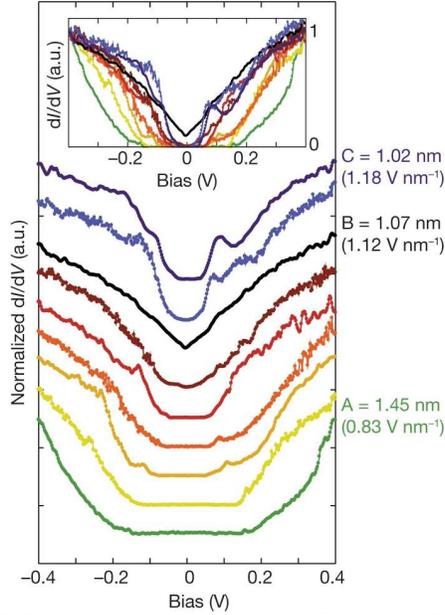

Fig. 2. Scanning tunneling spectroscopy of bilayer Na$_3$Bi at different tip-sample separations, corresponding to different electric fields. Tip-sample separation and electric fields are shown labelled for curves A, B, C. Adapted from [2].

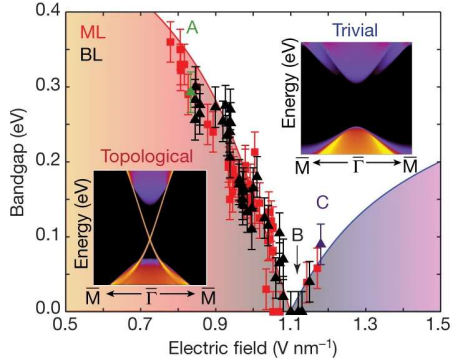

Fig. 3. Electric field dependence of the bandgap in monolayer (ML; red) and bilayer (BL; black) Na$_3$Bi. Adapted from [2].

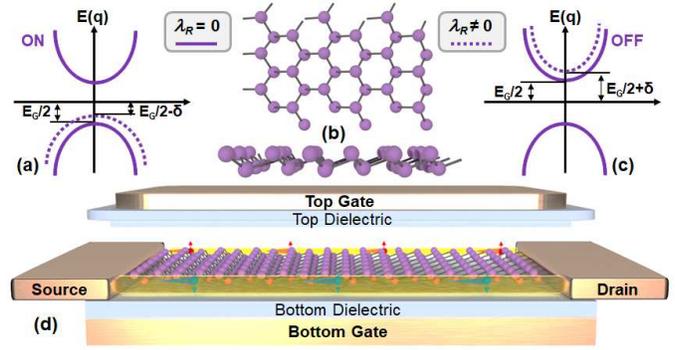

Fig. 4. Topological quantum field effect transistor. In the absence of a gate electric field, a QSH insulator hosts dissipationless helical conducting channels with a minimum value of the quantized conductance 2 $e^2/h$ (ON state of TQFET (a)). When the gate electric field exceeds a threshold limit, the thin QSH insulator layer (staggered honeycomb lattice (b)) enters into the trivial regime, in which the minimum value of the conductance drops to zero (OFF state of TQFET (c)). Such electric field switching is accompanied by the topological quantum field effect which enhances the topological phase transition driven by a gate electric field and reduces the subthreshold swing (a, c). Here, δ represents the shift in nontrivial (a) and trivial (c) band gap $E_G$ due to topological quantum field effect. (d) Geometry of topological quantum field effect transistor. Adapted from [Nad21].

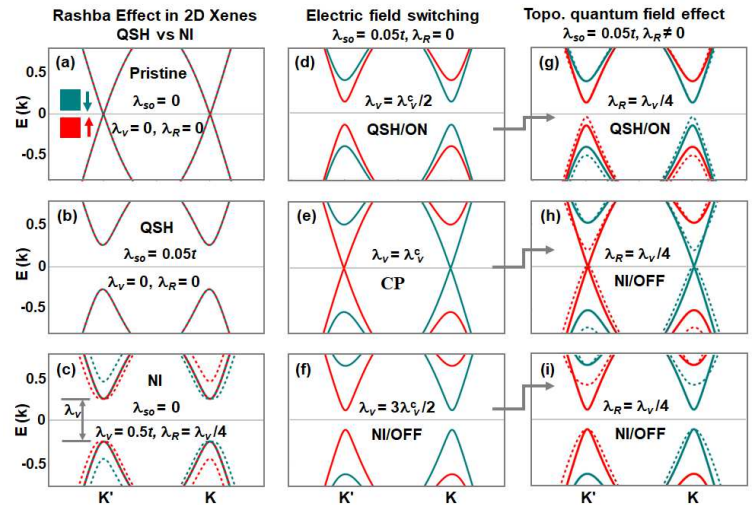

Fig. 5. Electric field switching and Topological quantum field effect. (a-c) Band dispersion for pristine Xene sheets with Dirac dispersion at valleys K(K$_0$) (a). Intrinsic SOI λ$_{so}$ opens nontrivial band gap and leads to QSH phase (b). In the purely gate-induced normal insulating (NI) phase (λ$_{so}$ = 0), where bands remain two-fold spin-degenerate, Rashba SOI lifts the spin-degeneracy but the trivial band gap remains insensitive to the Rashba effect (c). (d-f) In the absence of Rashba SOI, a uniform out-of-plane electric field drives QSH (d) to normal insulating phase (f) while passing through a critical point (λ$_v$ = λ$_v^c$) where valleys K(K$_0$) are perfectly spin-polarized hosting spin down (up) gapless states (e). (g-i) With Rashba SOI of strength λ$_R$ = λ$_v$/4 where λ$_v$ is same as in the respective panels of (d-f), topological quantum field effect reduces the nontrivial band gap (g), opens the trivial band gap at the threshold gate electric field (h), and enhances the trivial band gap (i). The solid (dashed) lines represent band dispersion in the absence (presence) of Rashba SOI. Adapted from [Nad21].

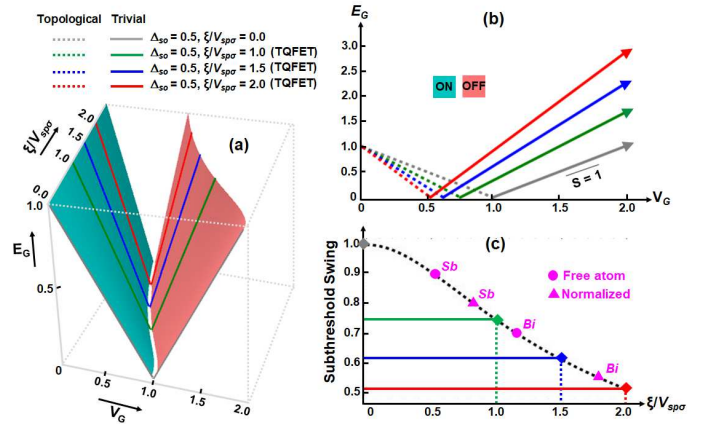

Fig. 6. Topological quantum field effect on band gap, threshold gate voltage, and subthreshold swing. The green, blue, and red lines represent the variation of band gap (a), threshold gate voltage (b), and subthreshold swing (c) corresponding to atomic SOI and Slater-Koster parameter ratio ζ/V$_{spσ}$ = 1, 1.5, 2 respectively which encodes the topological quantum field effect. (a) Nontrivial (trivial) bulk band gap E$_G$ decreases (increases) sharply with increasing ζ/V$_{spσ}$. Accordingly, the threshold gate voltage (b) and subthreshold swing (c) decreases with increasing ζ/V$_{spσ}$. Magenta circles (triangles) represent the subthreshold swing for TQFET based on antimonene and bismuthene with free atomic (normalized) SOI. Here we assume that $d_z \approx$ and sinθ ≈ 1 for quasi-planar/low-buckled honeycomb lattice. Adapted from [Nad21].

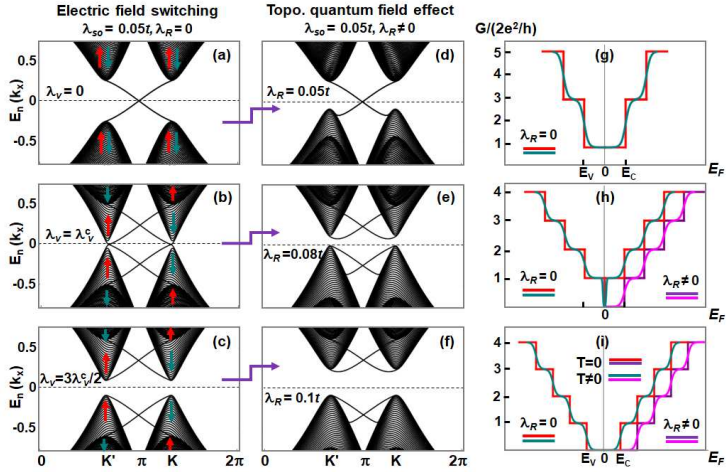

Fig. 7. Edge state dispersion and conductance quantization for semi-infinite honeycomb strip with zigzag edges. (a-c) Electric field switching via topological phase transition showing helical edge states as available conducting channels in the ON state (a), critical gapless phase (b) and the trivial insulator or OFF state (c). (d-f) Topological quantum field effect in the "ON" state reduces the band gap by shifting the valence band along energy axis (d). On the other hand, at the critical point (e) and in the "OFF" state (f), topological quantum field effect enhances the trivial band gap where the maximum of valence band remains pinned but the minimum of conduction band is lifted along energy axis. (g-i) Quantized conductance in terms of the number of modes $M(E)$ available at a given energy for TQFET in the QSH phase (g), at the critical point $\lambda_v = \lambda_v^c$ (h) and in the OFF state (i). While conduction band minimum remains pinned in ON state (g), (h) and (i) also show the shift of conduction band minimum in OFF state, along Fermi energy, induced by topological quantum field effect. In the ON (OFF) state, conductance jumps from 2 $e^2/h$ (0) to 6 $e^2/h$ (2 $e^2/h$) at the tip of conduction/valence bands lying at the energy $E_c$ ($E_v$) while the electric field switching (b, e, h) is based upon manipulating the minimal conductance between 2 $e^2/h$ and 0 and topological quantum field effect enhances this process. The spin chirality of edge states is same as shown for the tips of connecting Dirac cones. Adapted from [Nad21].

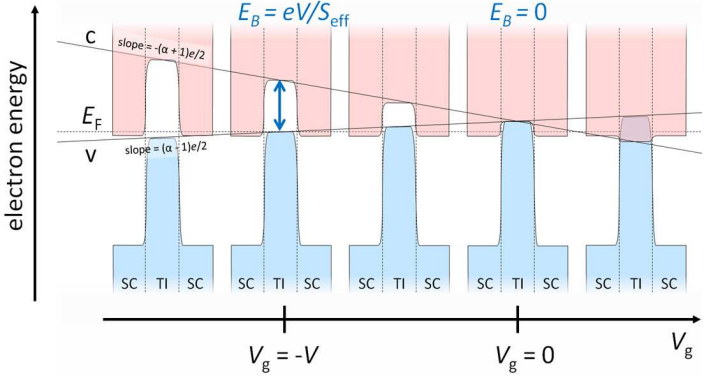

Fig. 8. Schematic band diagrams of the unipolar NC-TQFET as a function of gate voltage applied to a single gate $V_g$.

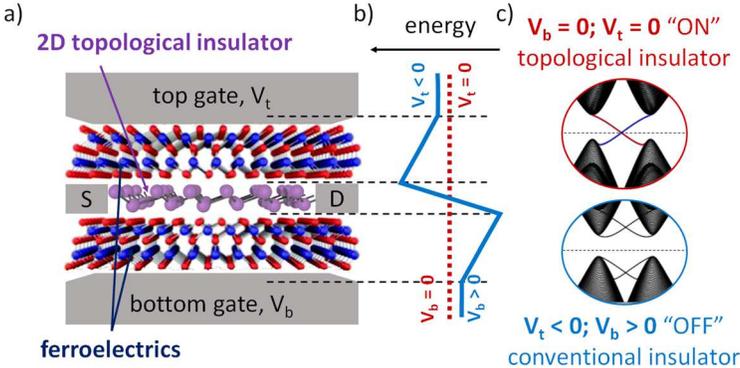

Fig. 9. Schematic of negative capacitance topological quantum field-effect transistor. (a) Structure of the device. (b) Electrostatic potential energy as a function of distance across the device in "on" (red) and "off" (blue) states. (c) Band diagrams in "on" (topological insulator, red) and "off" (conventional insulator, blue) states.

| | $\xi(eV)$ | | | $V_{sp\sigma}$ | $S^{*\ddagger}$ |
|---|---|---|---|---|---|
| | Free atom | Normalized | Experimental[†] | (eV) | $\Delta_R \neq 0$ |
| Graphene | 0.006 | 0.009 | - | 5.580 | 0.999[0.999] |
| Silicene | 0.028 | 0.044 | (-)0.044 | 2.54 | 0.999[0.999] |
| Germanene | 0.2 | 0.29 | (0.2)0.29 | 2.36 | 0.996[0.993] |
| Stanene | 0.6 | 0.8 | (0.48)0.77 | 1.953 | 0.961[0.934] |
| Arsenene | 0.29 (0.36) | 0.421 | - | 1.275 | 0.978[0.955] |
| Antimonene | 0.6 (0.8) | 0.973 | - | 1.170 | 0.904[0.802] |
| Bismuthene | 1.5 | 2.25 | - | 1.3 | 0.707[0.568] |

Table 1: Strength of atomic SOI ξ, Slater-Koster parameter $V_{sp\sigma}$ and sub-threshold swing S* for TQFET based on group-IV and V Xenes. Here we assume that $d_z \approx z$ and $\sin\theta \approx 1$ for quasi-planar/low-buckled honeycomb lattice. Similar to other group-IV and V elements, a normalization factor of 3/2 is assumed for bismuthene compared to the free atomic SOI. Adapted from [Nad21].

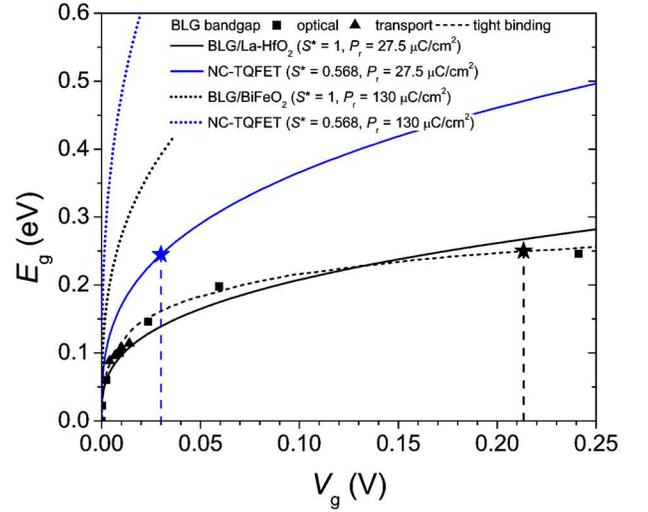

Fig. 10. Bandgap as a function of gate voltage for a bilayer graphene (BLG)-ferroelectric transistor (S* = 1), and a hypothetical NC-TQFET with similar properties but strong spin-orbit coupling (S* = 0.568). The experimentally measured electric-field-dependent bandgap values for BLG from optical spectroscopy[8] and electronic transport[9] are also shown, along with a tight-binding calculation[8]. The remanent polarization $P_r$ = 27.5 μC/cm$^2$ corresponds to La-doped HfO$_2$ [6], and $P_r$ = 130 μC/cm$^2$ corresponds to tetragonal-like BiFeO$_3$ [12].